\documentclass{article}
\usepackage{graphicx}
\usepackage{mathtools}
\usepackage{amsmath}
\usepackage{amssymb}
\usepackage{dsfont}
\usepackage{bbold}
\usepackage[a4paper, left=2cm, right=2cm, top=2cm, bottom=2cm]{geometry}
\usepackage{caption}
\usepackage{subcaption}
\usepackage{float}
\usepackage[normalem]{ulem}
\usepackage{microtype}
\usepackage{xargs}
\usepackage{xcolor}
\usepackage[colorinlistoftodos, prependcaption]{todonotes}
\usepackage{multirow}
\usepackage{booktabs}

\usepackage[backend=biber, sorting=none, citestyle=numeric-comp, bibstyle=ieee, url=false, doi=false, isbn=false, eprint=false]{biblatex}
\AtEveryBibitem{%
  \clearfield{number} % Remove number
  \clearfield{note} % Remove publisher
  \clearfield{address} % Remove address
  \clearfield{month}
  \clearfield{day}
  \clearfield{doi}
}
\title{
Localization in tight-binding models with power-law distributed couplings
}

\author{Maximilian Weigmann, Luca Schaefer, and Barbara Drossel}
\date{Institute for Condensed Matter Physics, Technical University of Darmstadt,
 Hochschulstraße 6, 64289 Darmstadt, Germany}

\begin{document}
\maketitle
\abstract{
We study the localization properties of 1D and 2D tight-binding models with power-law distributed couplings by comparing the spectrum and the localization properties of the eigenmodes of the Laplacian and the adjacency matrix, using numerical diagonalization of these matrices for different system sizes and connectivities. These two matrices are relevant for different types of dynamical processes. While all eigenmodes of the adjacency matrix are localized for sufficiently large system sizes, the Laplacian matrix always leads to a small proportion of system-spanning modes due to a conservation law, and therefore to power-law tails in the probability distribution of the participation ratio and its relation to the eigenvalues. In one dimension, the exponent of these power laws change continuously with the exponent that characterizes the distribution of couplings. In two dimensions, the modes with the largest relaxation times change from system-spanning to localized when the exponent of the distribution of couplings becomes larger than 0.75. We provide phenomenological explanations for all these findings. 
}

\section{Introduction}
The transport of charges in a semiconductor, the propagation of light in a crystal, and the dispersal of species in a patchy landscape are all affected by the presence of disorder. 
Since the seminal work by Anderson in 1958~\cite{Anderson.1958}, it is well known that a sufficiently strong disorder leads to a localization of eigenvectors of such systems~\cite{kramer_localization_1993}.
Anderson considered a tight-binding model for electrons with random on-site potential. Subsequent work generalized these findings to tight-binding models with disorder also in the hopping terms, and to other wave equations that can be mapped on the same type of eigenvalue problem~\cite{kramer_localization_1993,eilmes_two-dimensional_1998,deanVibrationsGlasslikeDisordered1964,dominguez-adameDelocalizedVibrationsClassical1993}. 
Analytical calculations have shown that all eigenmodes are localized in one dimension~\cite{delyonOnedimensionalWaveEquations1983}, and that two is the critical dimension above which a  phase transition from localization to delocalization occurs as the strength of disorder is reduced~\cite{abrahamsScalingTheoryLocalization1979}. 

However, some types of models have extended modes even in one and two dimensions. One example is the random-coupling model, which has only bond disorder and no site disorder. The Hamiltonian of this system has the form of an adjacency matrix. Such models are, for instance, relevant for epidemic spreading processes~\cite{pastorsatorrasEpidemicProcessesComplex2015a}. Due to the bipartite structure of the lattice, this model has a symmetry which leads to an extended mode exactly at the band center \cite{eilmes_two-dimensional_1998}. Another example is the diffusion model, for which  site and bond disorder are correlated such that the sum of all entries in a row vanishes.  The Hamiltonian of this system is a Laplacian matrix, and it has a conserved quantity, leading to system-spanning eigenvectors in finite systems even in the presence of strong disorder and even in one dimension~\cite{hirotsugamatsudaLocalizationNormalModes1970,schaeferScalingBehaviourLocalised2024}. Such diffusion models apply, for instance, to classical harmonic chains and to the diffusion of a substance or a biological population through a set of coupled sites~\cite{dattaAbsenceLocalizationOnedimensional1994,dominguez-adameDelocalizedVibrationsClassical1993,bernasconiClassicalDiffusionOneDimensional1978,brechtelMasterStabilityFunctions2018}. For one- and two-dimensional systems, it was shown that the number of system-spanning eigenvectors scales as $\sqrt{N}$ with $N$ being the number of sites~\cite{hirotsugamatsudaLocalizationNormalModes1970, dominguez-adameDelocalizedVibrationsClassical1993,paytonDynamicsDistortedHarmonic1967,p.deanVibrationsTwodimensionalDisordered,schaeferScalingBehaviourLocalised2024}.
In the limit of infinitely large systems, this is a vanishing fraction of all modes.

The essential features of these models should not depend on the precise shape of the probability distribution of the coupling strengths, unless many couplings 
have values close to zero, changing fundamentally the propagation processes in these systems. In one dimension, it has been shown that the power laws of the 
diffusion model change their values when the interval from which the coupling strengths are chosen extends to zero \cite{Schaefer.2025}. For a power-law distribution of couplings, anomalous diffusion was found in \cite{Alexander1981}.
In two dimensions, a percolation transition occurs when the proportion of zero couplings exceeds the threshold for bond percolation, which is 0.5. Above the percolation threshold, we expect universal behavior, as a coarse-grained view that averages over subsystems of many lattice sites always gives a Gaussian distribution of effective couplings. 

Since two is the critical dimension of the localization transition, two-dimensional systems show strong finite-size effects, making it difficult to see the true asymptotic behavior in numerical evaluations of the eigenmodes.  However, increasing the strength of disorder increases localization and reduces finite-size effects. This was our main motivation to investigate the model with power-law distributed link strengths $f(t) \sim t^{-s}$. As the exponent $s$ approaches 1 from below, the proportion of links within a small distance from 0 increases ever more. Such power-law distributions are scale-free and occur frequently in nature, which is an additional reason to study the model version with power-law distributed couplings. Apart from stronger localization we find in the one-dimensional diffusion model nonuniversal power laws for the density of states and for the dependence of the participation ratio of a mode on the associated eigenvalue. In two dimensions, the modes with the smallest eigenvalues (or, equivalently, the largest relaxation times) change from system-spanning to localized on a few lattice sites when the exponent $s$ exceeds 0.75. 
In the following, we will not only present the results of our numerical evaluation of the eigenmodes, but will also provide phenomenological explanations for the various findings.

\section{Models}

The starting point of our investigation is the discrete tight-binding Schrödinger equation
\begin{align}\label{eq:schroedinger_equation}
    E\psi_n = \epsilon_n\psi_n - \sum_m t_{nm}\psi_{m} 
\end{align}
where the sum is taken over all sites $m$ that are nearest neighbors of site $n$. 
This equation is obtained from the tight-binding model, which describes non-interacting electrons in a lattice, with the attraction between lattice ions and electrons being so strong that electrons are localised in the vicinity of ions. This equation can also be obtained directly by discretising the one-dimensional Schr\"odinger equation. The $t_{nm}$ are the hopping matrix elements which result from the kinetic energy term and characterise the transition amplitudes of an electron from one potential well to another. These and the on-site energies $\epsilon_n$ may be random variables in a system with disorder.

We consider the case of power-law distributed couplings $t_{nm}$,
\begin{align}
    f(t) = \begin{cases}
(1-s) t^{-s}\text{ for } 0<t<1, \\
0 \text{ \qquad \quad\,\,\, otherwise }
\end{cases} ~.
\end{align}
The parameter $s \in(0,1)$ determines the strength of disorder. For $s \to 0$, the power law approaches the constant distribution that is most often chosen in the literature. 

We distinguish two different models depending on the relation between the couplings $t_{nm}$ and  the on-site energies $\epsilon_n$. The  random-coupling model (RCM) has $\epsilon_n=0$, as for instance done in \cite{soukoulisOffdiagonalDisorderOnedimensional1981,eilmes_two-dimensional_1998}. The Hamiltonian of this model is identical to the weighted adjacency matrix associated with the system. This model has a symmetry since the lattice is bipartite: to each mode with energy $E$ and eigenvector $\{\psi_n\}$ there is an equivalent mode with energy $-E$ and eigenvector $\{(-1)^n \psi_n\}$, with $(-1)^n$ being $1$ on one sublattice and $-1$ on the other. Because of this symmetry, we show in all figures only the modes with positive energy. 

The diffusion model (DM) has $\epsilon_n = \sum_m t_{nm}$, and has been studied, for instance, in \cite{bernasconiClassicalDiffusionOneDimensional1978,pinskiAndersonUniversalityModel2012}. The time-dependent version of the DM (with $i\hbar \partial\psi_n/\partial t$ on the left-hand side instead of $E\psi_n$) has a conserved quantity $\sum_n \psi_n$. While it may be difficult to find a motivation for this model within the framework of the quantum-mechanical description of electrons in disordered lattices, there are various applications of this model in other fields of physics. In classical mechanics, the model describes a system of identical masses $m$ coupled by springs,
\begin{equation}
    m\ddot x_n = -\sum_l D_{nl}(x_{n}-x_{l}) \, ,
\end{equation}
where the $x_n$ represent the deviations from the equilibrium positions and $D_{nl}$ are the spring constants between the masses at sites $n$ and $l$. This system of equations can be solved with the ansatz $x_n(t)=x_n(0)e^{-i\omega t}$, leading to 
\begin{equation}
    m_n \omega^2 x_n = \sum_l(D_{nl}(x_{n}-x_{l})\, .
\end{equation}
This is an equation for the oscillating eigenmodes of the system, with $\omega$ being the oscillation frequency. 

Another, very broad, field of applications of the DM is that of diffusion processes between discrete sites (hence our name for the model). In this case, the $\psi_n$ are replaced by the amounts $c_n$ of the substance that diffuses between sites, which may be, for instance, molecules, heat, or individuals of a population, depending on the application. In ecological applications, the sites represent habitats. The time evolution of such systems is described by the equations
\begin{equation}
    \frac{\partial c_n}{\partial t} = -\sum_{l}D_{nl}(c_n-c_l) \label{DMt}
\end{equation}
with the sum being taken over the nearest neighbors of $n$. The $D_{nl}$ are now the rates of transfer between sites or habitats $n$ and $l$. The total amount $\sum_n c_n$ is a conserved quantity. With the ansatz $c_n(t)=c_n(0)e^{-t/\tau}+ c_n^{\mathrm{eq}}$, one obtains an equation for the relaxing eigenmodes,
\begin{equation}
    \frac{c_n}{\tau} = \sum_{l} D_{nl}(c_n-c_l)\, , \label{DM2}
\end{equation}
with the relaxation constant $1/\tau$. 

The Hamiltonian of the diffusion model is identical to the discretized Laplacian matrix of the system.

 \section{Methods}
In order to quantify the localization behavior of the models, we calculate the eigenvectors and associated energies of Eq.~\eqref{eq:schroedinger_equation} by diagonalizing the Hamiltonian using
the Julia library \texttt{LinearAlgebra}. We then determine the participation ratio $P$ of each eigenvector and evaluate the probability distribution of $P$ and the relation between $P$ and the energy $E$. 

The participation ratio \cite{kramerLocalizationTheoryExperiment1993, eilmesTwodimensionalAndersonModel1998} of the $i$-th eigenstate $\psi_i$ is defined as
\begin{align}
    P_i=\frac{\left(\sum_n^N|\psi_{i,n}|^2\right)^2}{\sum_n^N|\psi_{i,n}|^4}~,
\end{align}
where the numerator is 1 when the wave function is normalized.
This definition gives $P=1$ for a state that is fully localized on one site, and  $P=N$ for a state with the same amplitude at every site. A special case is the participation ratio of the sine-shaped eigenmodes of the system without disorder, for which we have  $P=2N/3$ in one dimension and $P=4N/9$ in two dimensions. 

In order to have comparable energy scales for different power-law exponents $s$, we normalize all energies (i.e. eigenvalues) by dividing by the average hopping element, 
\begin{align}
    E^{'} = \frac{E}{\langle t \rangle}\, .
\end{align}

The system sizes and ensemble sizes used in our numerical calculations are summarized in Table \ref{tab:Ensemble_Sizes}. We used smaller sample sizes for larger systems, to keep the total number of eigenmodes approximately constant. 
\begin{table}[H]
    \centering
    \caption{System sizes and ensemble sizes used in 1D and 2D.}
    \begin{tabular}{c|ccccc|ccccc}
        \bfseries Dimension&\multicolumn{5}{c|}{\bfseries 1D} & \multicolumn{5}{c}{\bfseries 2D} \\ [5pt]
        \hline
         System size $N$ & 640 &1280&2560&5120&10240&625& 1444&3249&7056&15625\\[5pt]
        \hline
         Ensemble size & 950 & 800 & 400 & 200& 100&990&990&480&221&100 \\ [5pt] 
    \end{tabular} 
    \label{tab:Ensemble_Sizes}
\end{table}
\section{Results}
\subsection{The diffusion model in one dimension}
\begin{figure}[H] 
    \centering
    \includegraphics[scale=0.35]{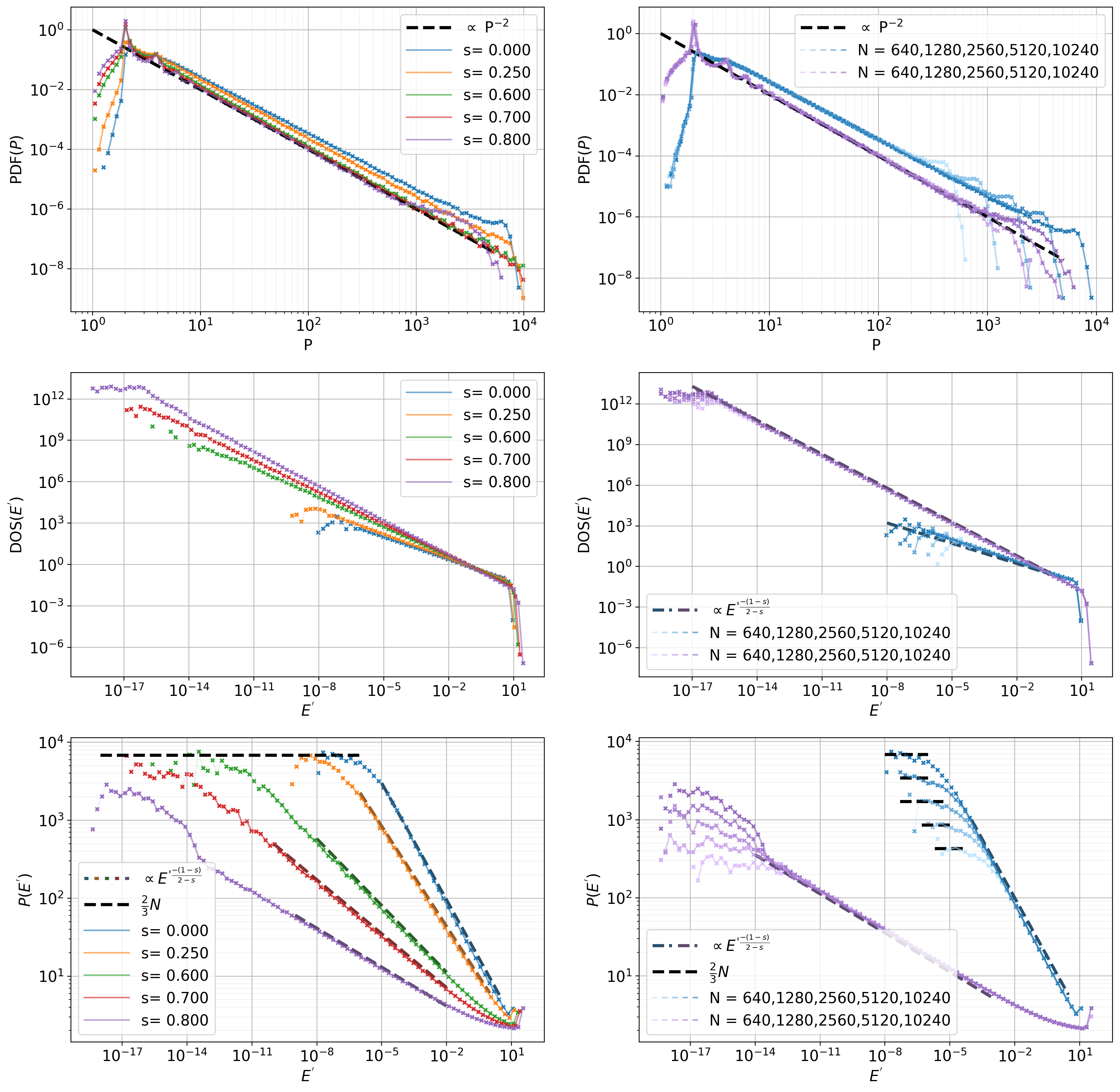}
    \caption{Probability distribution of the participation ratio (top), density of states (middle), and relation between participation ratio and eigenvalue (bottom) for the 1D diffusion model. The left graphs show the data for different power law exponents $s$, the right graphs show the data for different system sizes, for a large ($s=0.8$) and a small value ($s=0.0$) of $s$. The tilted dashed lines indicate power laws obtained from theoretical considerations; the horizontal dashed lines indicate the participation ratio of sine functions, which prevail for large wavelengths.} \label{Uebersicht/OVV_DM_tav_4_1D}
\end{figure}

For the one-dimensional DM, the probability distribution of the participation ratio  is a power law with exponent $-2$ irrespective of the value of the exponent $s$ of the distribution of couplings (see Figure~\ref{Uebersicht/OVV_DM_tav_4_1D}, first line). This power law can be obtained by the following simple argument: We assume that the participation ratio of an eigenmode is a measure of its spatial extension, and that the boundaries of the eigenmode are given by the weakest couplings over its extension. Given that one of these boundaries has a coupling strength $t$, the average distance to a coupling that is weaker or equal is of the order 
\begin{equation}
   P \sim  \left[\int_0^t \tau^{-s} \mathrm{d}\tau\right]^{-1} \sim t^{s-1}\,. 
\end{equation}
This allows us to calculate the probability distribution of $P$ from that of $t$, resulting in  $\mathrm{PDF}(P) \sim t^{-s} \mathrm{d}t/\mathrm{d}P \sim P^{-2}$. This estimate is better when the differences in coupling strengths are more pronounced, i.e. when $s$ is closer to 1, just as our data show.  Changing the system size changes the cutoff, but not the slope of our data (see top right in Figure \ref{Uebersicht/OVV_DM_tav_4_1D}). 
The density of states and the relation between the participation ratio $P$ of an eigenmode and its energy $E$ follow also power laws, but now the exponents change with $s$ (see middle and bottom line of Figure \ref{Uebersicht/OVV_DM_tav_4_1D}). Let us first consider the relation between $P$ and $E$ (bottom graphs): The slope is less steep when $s$ is larger, and its value can be obtained by the following consideration: In the DM, the inverse of the "energy" $E$ of an eigenmode is the time it needs to relax. During this relaxation, amplitude must be transported over a distance of the order of the extension $P$ of the mode, namely from the region with positive amplitude to the region with negative amplitude. This transport occurs via a diffusion process. The average transport time through one link is given by $\langle 1/t \rangle_P \sim t_{\mathrm{min},P}^{-s}$, with 
$t_{\mathrm{min},P}$ being the smallest coupling value occurring over the distance $P$, which is 
$\sim P^{-1/(1-s)}$. The time for diffusive transport over the distance $P$ is therefore given by 
\begin{equation}
    E^{-1} \sim P^2 \langle 1/t \rangle_P\sim P^{(2-s)/(1-s)} \text{ or }  P \sim E^{-(1-s)/(2-s)}\, ,
\end{equation}
in excellent agreement with our simulation data. This anomalous diffusion has already been described in \cite{Alexander1981}. Again, changes in system size within the considered range affect the cutoff but not the slope. The density of states $\mathrm{DOS}(E)$ is not an independent function but follows from the other two via the  relation 
\begin{equation}
    \mathrm{DOS}(E) = \mathrm{PDF}(P) \mathrm{d}P/\mathrm{d}E \sim E^{-1/(2-s)}\, .
\end{equation}
For the case $s=0$, i.e. a constant coupling distribution, the density of states is given by the inverse square root law of the tight-binding model without disorder. This means that for small $E$, where the eigenmodes span a large distance, these eigenmodes resemble the sine waves of the model without disorder, as has been found before in \cite{Schaefer.2025}. The participation ratio associated with a sine wave is $2N/3$, which agrees well with the maximum $P$ value seen in the bottom row of Figure \ref{Uebersicht/OVV_DM_tav_4_1D} for $s=0$. 
 
In addition to these power laws, the data also show that relaxation times become longer (i.e. the eigenvalues $E$ become smaller) and the participation ratios become smaller when $s$ is larger, i.e. when there are more weak couplings in the system. Both trends can be understood intuitively: A larger number of small couplings increases the time for transporting amplitude through a link, and the smallest couplings determine the longest time scales. Furthermore, many small couplings mean larger disorder  and larger differences in the amplitudes at different sites, which in turn leads to smaller $P$ values for modes that have a similar overall extension. For the larger $s$ values, the probability distribution of $P$ shows a pronounced peak at $P=2$. An inspection of the eigenstates associated with these $P$ values shows that the states are dominated by two neighboring sites with opposite and equal amplitude, delimited by two weak couplings that are much smaller than the coupling connecting the two sites. Relaxation of such a mode occurs mainly by flux through the coupling connecting the two sites. 

\subsection{The random-coupling model in one dimension}

\begin{figure}[H] 
    \centering
    \includegraphics[scale=0.35]{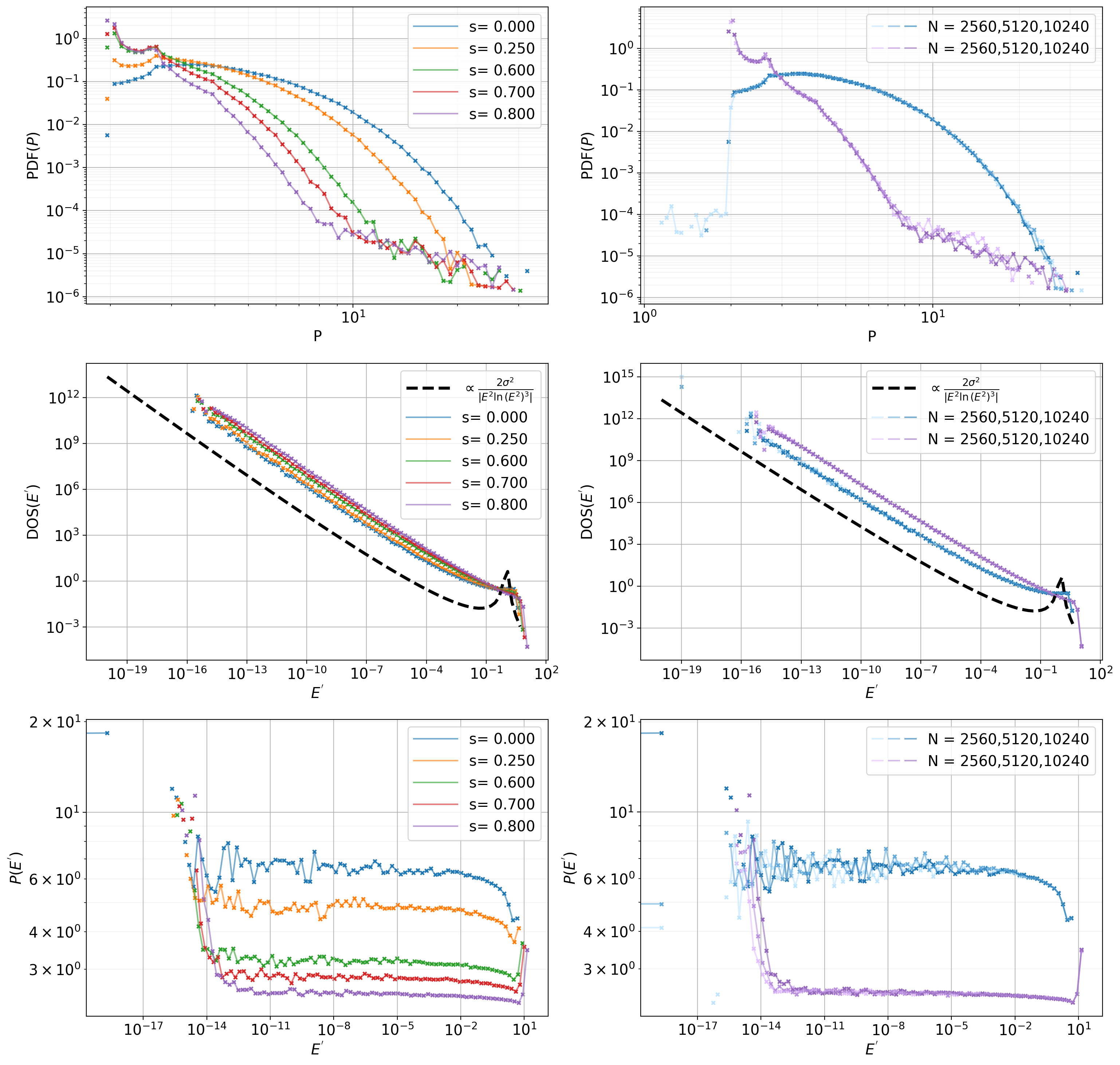}
    \caption{Probability distribution of the participation ratio (top), density of states (middle), and relation between participation ratios and eigenvalues (bottom) for the 1D random-coupling model. The graphs on the left show the data for different power law exponents $s$, the graphs on the right show the data for different system sizes, for a large ($s=0.8$) and a small value ($s=0.0$) of $s$. The dashed line represents the Dyson singularity $\sim (E \ln^3 E)^{-1}$.} \label{Uebersicht/OVV_RCM_tav_2_1D}
\end{figure}

The one-dimensional RCM shows full localization of all eigenmodes, with the probability distribution of the participation ratio becoming independent of the system size when the system size is large enough (see first row of Figure \ref{Uebersicht/OVV_RCM_tav_2_1D}). 
For $s$ close to 1, there occurs a flatter tail at large $P$, which is a numerical artifact since small eigenvalues cannot be numerically distinguished, and therefore the algorithm yields eigenvectors that are linear combinations of smaller eigenvalues that all have the same energy within numerical accuracy. This feature is also visible in the bottom row, where $P$ versus $E$ is plotted. The density of states (middle row of Figure \ref{Uebersicht/OVV_RCM_tav_2_1D}) shows the Dyson singularity $(\sim E \ln^3 E)^{-1}$ discussed in \cite{dyson1953dynamics,eggarter1978singular}.

Just as for the DM, the typical participation ratios become smaller when $s$ increases towards 1. 

\subsection{The diffusion model in two dimensions}
In contrast to one-dimensional systems, in a two-dimensional system transport over large distances is not determined by the weakest couplings since there exist many alternative paths. The long-wavelength modes see an average coupling strength and should therefore in many ways be similar to those of a system without disorder. The exponent $s$ characterizing the distribution of couplings should lead to larger differences in the amplitudes between neighboring sites and therefore to smaller participation ratios. The largest relaxation times (i.e. smallest $E$) should thus become longer. 

\begin{figure}[H] 
    \centering
    \includegraphics[scale=0.45]{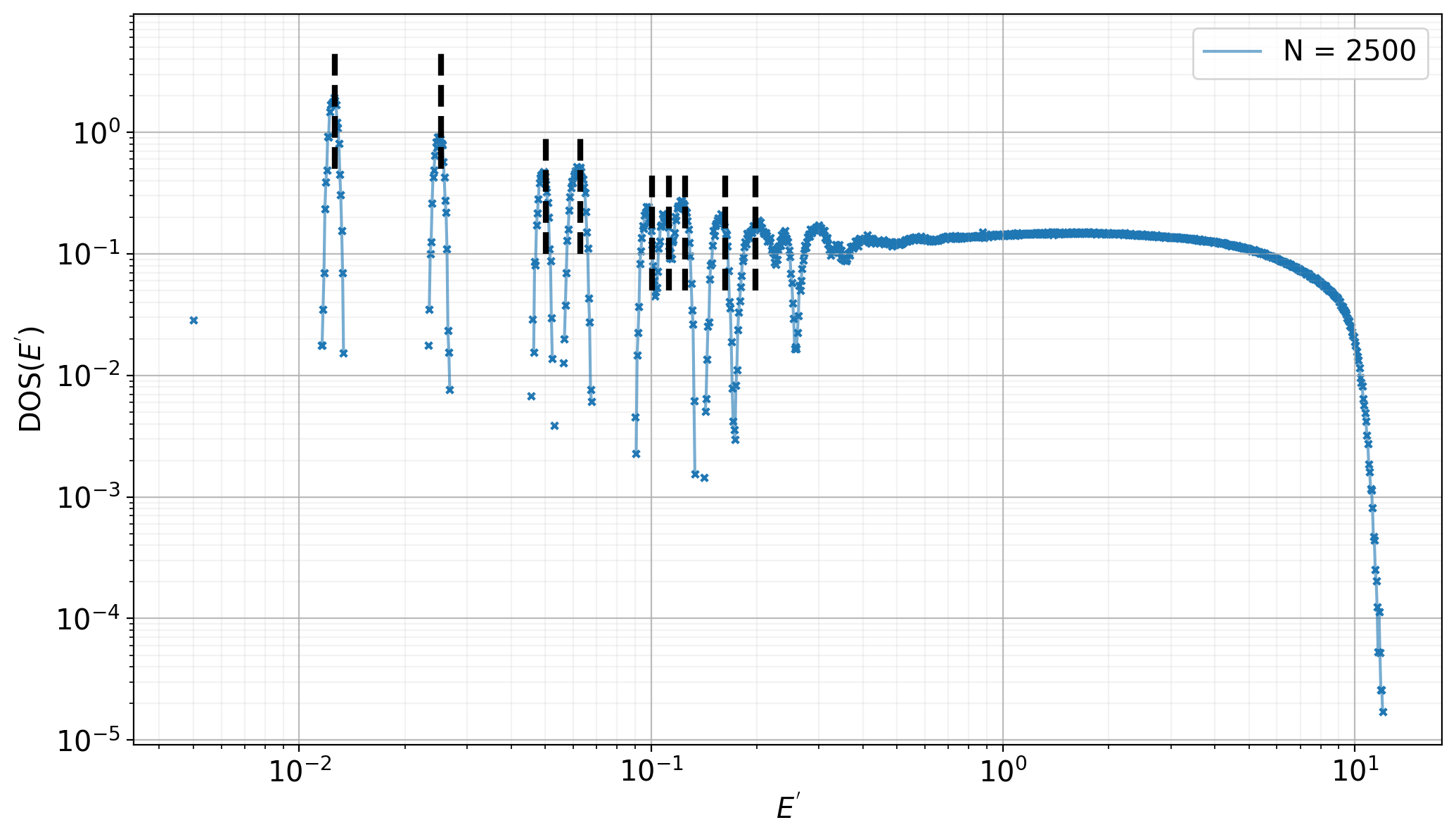}
    \caption{Density of states of the 2D diffusion model for $s=0$ and $N=2500$. The vertical dashed lines indicate the lowest eigenvalues of the system without disorder, where the eigenmodes are products of sine functions.} \label{Uebersicht/DOS_DM_examp_xy_0}
\end{figure}

\begin{figure}[H] 
    \centering
    \includegraphics[scale=0.35]{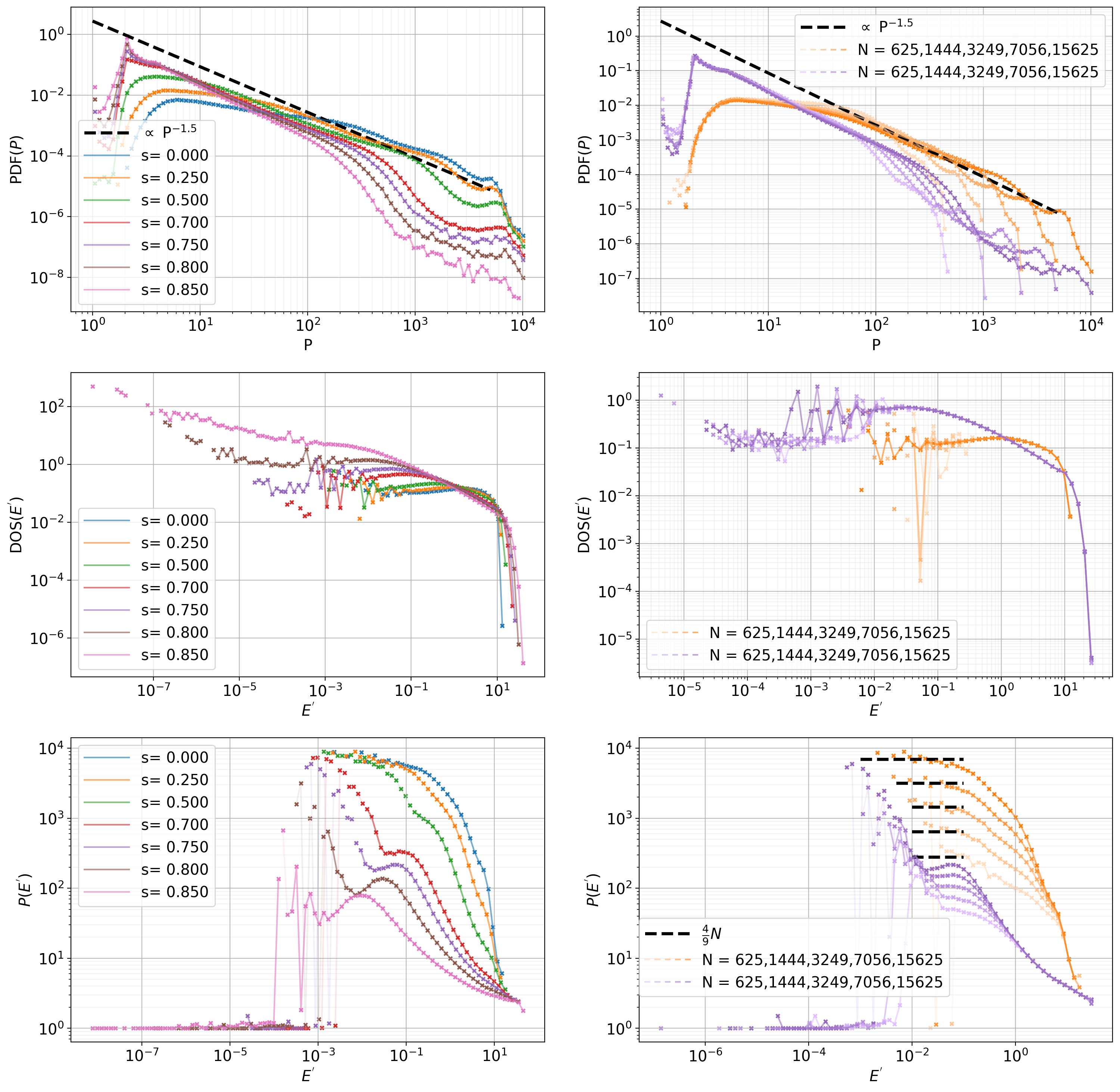}
    \caption{Probability distribution of the participation ratio (top), density of states (middle), and relation between participation ratio and eigenvalue (bottom) for the 2D diffusion model. The left graphs show the data for different power law exponents $s$, the right graphs show the data for different system sizes, for a large and a small values of $s$. The dashed lines in the top row show the theoretically predicted exponent $-1.5$.} \label{Uebersicht/OVV_DM_tav_4_2D}
\end{figure}

Our simulation data confirm these expectations and yield some additional insights. The density of states plotted in Figure \ref{Uebersicht/DOS_DM_examp_xy_0} for the case $s=0$ shows the flat behavior $\mathrm{DOS}(E) \propto \mathrm{const.}$ of the disorder-free two-dimensional tight-binding model. For the smallest energies, where the eigenmodes are sine waves with only a few wavelengths in both dimensions, the density of states shows discrete peaks close to the eigenvalues of the disorder-free system (indicated with vertical dashed lines). For $s$ larger than zero, the density of states also has a flat part at lower energies, but shows additionally a power-law like section that is larger for larger $s$, see Figure \ref{Uebersicht/OVV_DM_tav_4_2D}, middle row. The large-wavelength system-spanning modes have the participation ratios $2N/3$ and $4N/9$ associated with sine waves, and this is the cutoff value of $P$ seen in the top and bottom row of Figure \ref{Uebersicht/OVV_DM_tav_4_2D}. 

The probability distribution of the participation ratio appears to approach a power law $\sim P^{-1.5}$ when the system size increases, see Figure \ref{Uebersicht/OVV_DM_tav_4_2D}, top row. Due to strong finite-size effects, this power law is not really visible for $s=0$, but becomes clearer for larger $s$. The exponent $-1.5$ is expected on theoretical grounds, see \cite{Schaefer.2025,schaefer2026different}: For small energies, the eigenmodes extend over many lattice sites and are in each dimension similar to sine waves, with the amplitudes performing a random walk due to the disorder. The relative strength of disorder over the wavelength $\lambda$ is of the order $1/\sqrt \lambda$ due to the averaging over a distance $\lambda$. This means that the amplitude can decrease to zero via a random walk that covers $(\sqrt \lambda)^2=\lambda$ wavelengths. The value of $P$ associated with such a mode is $\sim \lambda^4 \sim E^{-2}$. Together with the constant density of states, this leads to an exponent $-1.5$ for the probability distribution of $P$ values.  

\begin{figure}[H] 
    \centering
    \includegraphics[scale=0.175]{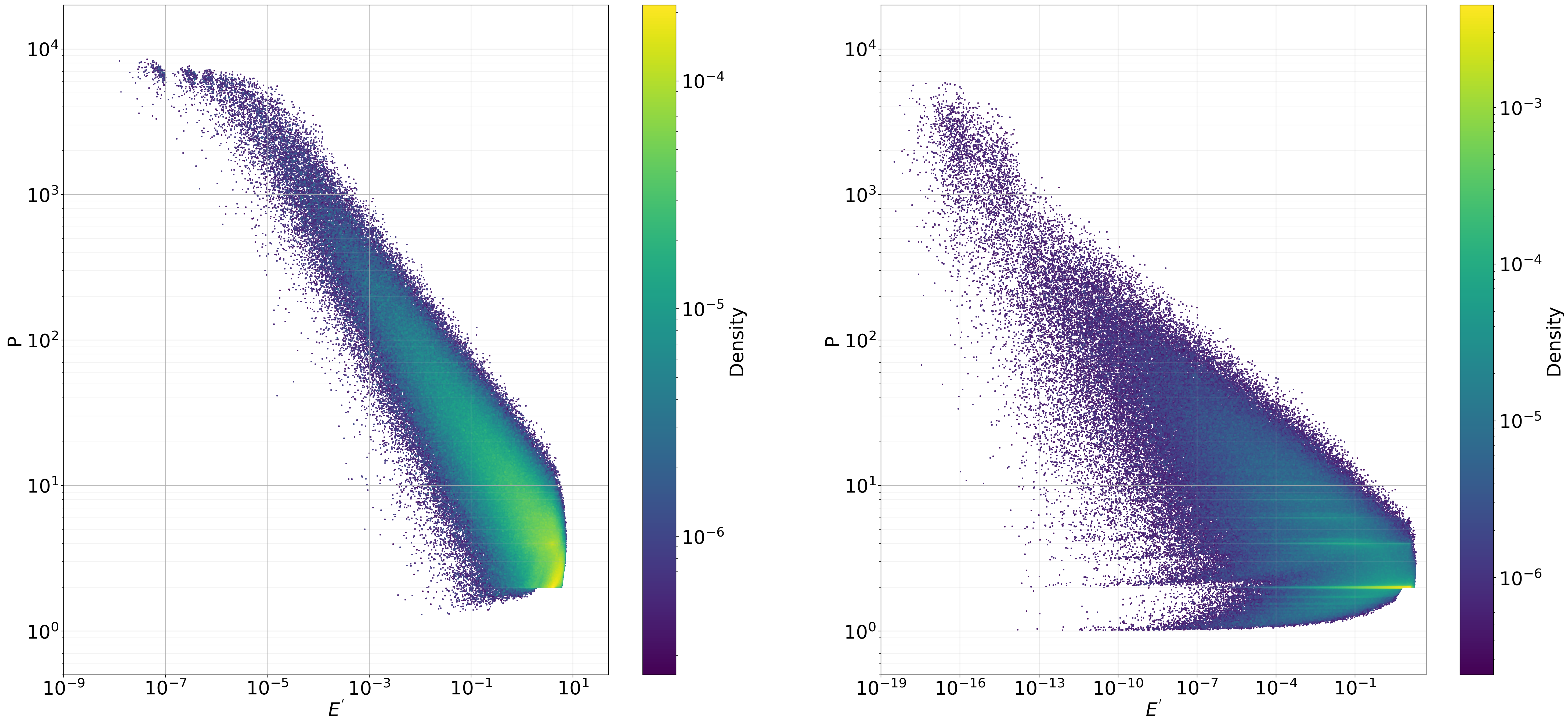}
    \caption{Each eigenmode of the 2D diffusion model is represented by a dot in the $P$-$E$ plane, with the color indicating the density of dots. The system size is $N=15625$, the power-law exponent is $s=0.0$ on the left and $s=0.85$ on the right.} \label{Uebersicht/E_Scatter_DM_tav_0_4_7_4_d}
\end{figure}

The relation $P \sim E^{-2}$ is not really visible in our data when $s$ is not small, see bottom row of Figure \ref{Uebersicht/OVV_DM_tav_4_2D}. The reason is that we plot the average value of $P$ associated with the respective $E$ values, with small $P$ contributing more and more with increasing $s$. This is illustrated in Figure \ref{Uebersicht/E_Scatter_DM_tav_0_4_7_4_d}, where each eigenmode of the ensemble is shown as a dot in the $P-E'$ plane for a small and a large value of $s$. A rough approximation to the $P$ versus $E$ data is given by the power law $P \sim E^{(4-4s)/(2-s)}$, which is an interpolation between the limits $P \sim E^{-2}$ for $s=0$ and $P \sim \mathrm{const.}$ for $s=1$.

Figure \ref{Uebersicht/E_Scatter_DM_tav_0_4_7_4_d} shows that for larger $s$ and at larger energies there are many eigenmodes with $P \simeq 2$. These are modes that are localized on a pair of sites that have amplitudes of opposite signs. They are connected by a strong coupling that determines the relaxation time, and are surrounded by weak couplings, which become numerous when $s$ is close to 1. We also see many eigenmodes with $P \simeq 1$ at very small energies when $s$ is larger than 0.75. These are modes that localize on a site that is surrounded by four weak links. 

\begin{figure}[H] 
    \centering
    \includegraphics[scale=0.45]{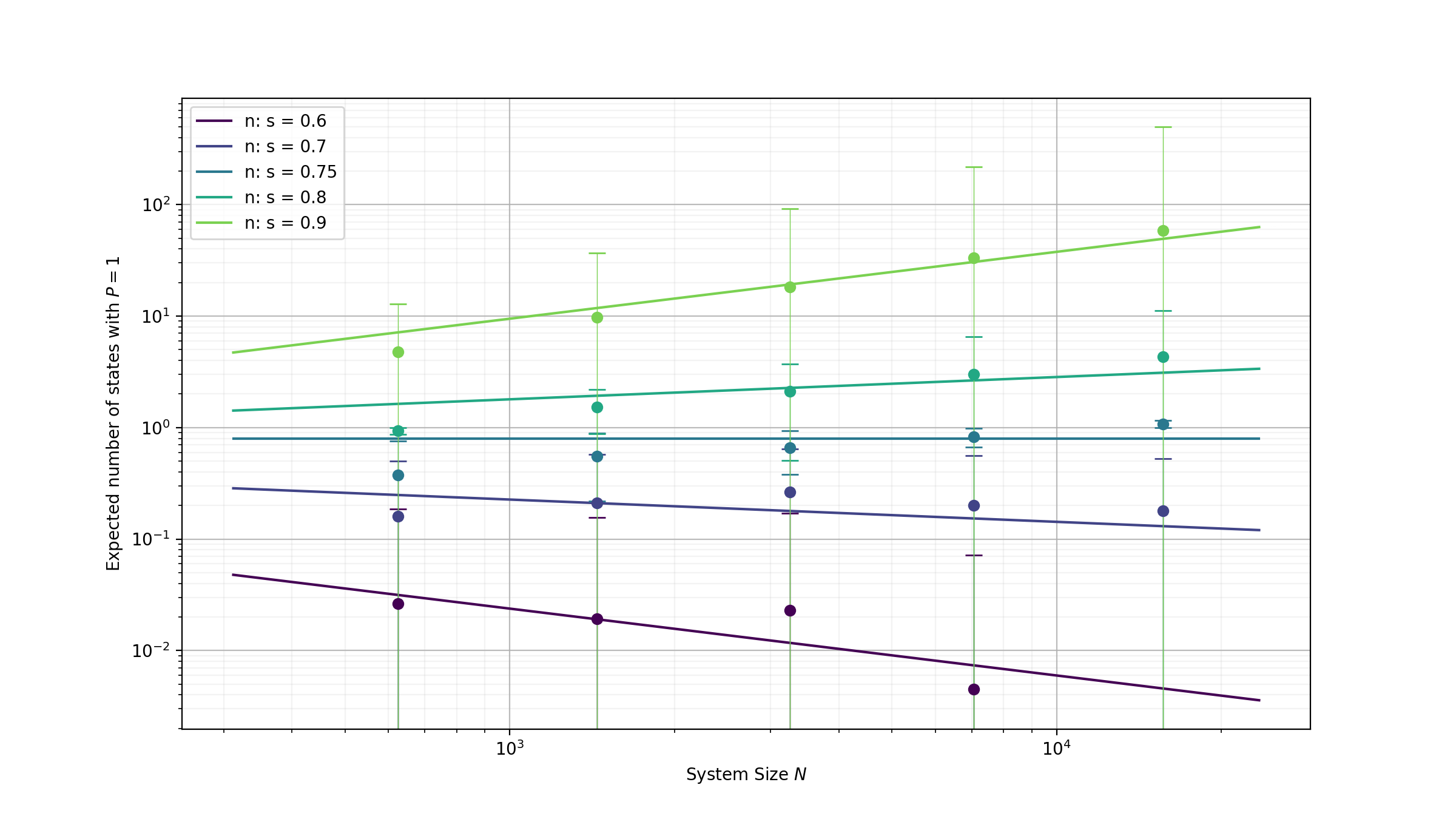}
    \caption{The average number of eigenmodes with $P=1$ as function of system size in the 2D diffusion model, for different values of $s$. The straight lines are the theoretical prediction $\sim N^{4s-3}$.} \label{Uebersicht/Expected_P}
\end{figure}

A quick estimate shows that the longest time scales (i.e. smallest $E$) become dominated by these $P \simeq 1$ modes when $s$ is larger than 0.75: 
The probability that all four couplings of a site are smaller than a value $t_0$ is $p_4 \sim t_0^{4-4s}$. On the other hand, a long-wavelength mode has an energy $E \sim \lambda^{-2} \sim 1/N$. The average number of $P \simeq 1$ modes that have an energy smaller than this is obtained by setting $t_0 = E \sim 1/N$, resulting in $Np_4 \sim N^{4s-3}$. This increases with increasing $N$ when $s$ is larger than 3/4. 
Figure  \ref{Uebersicht/Expected_P} confirms that this is indeed the case. 

\section{Random-coupling model in two dimensions}
\begin{figure}[H] 
    \centering
    \includegraphics[scale=0.35]{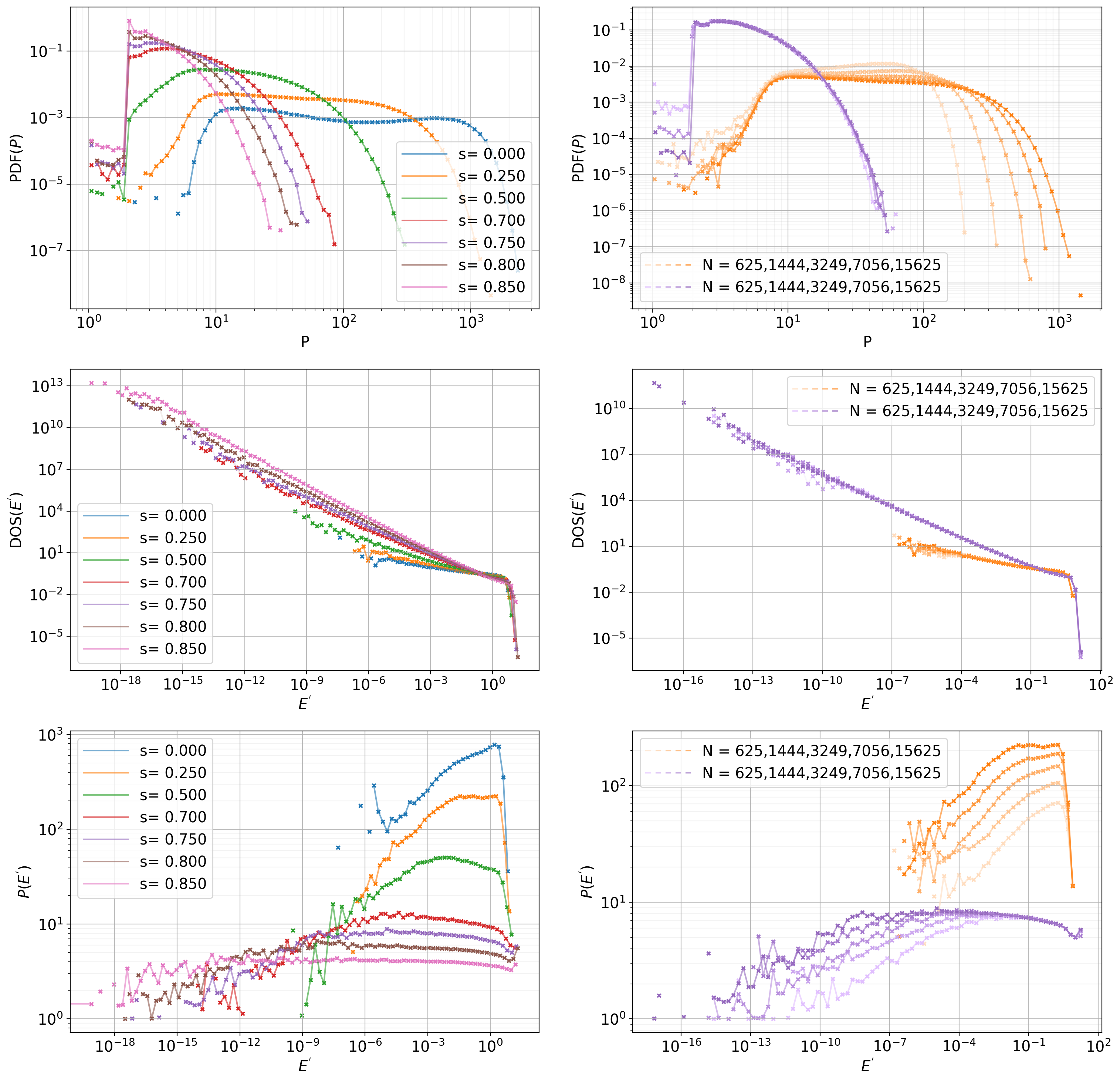}
    \caption{Probability distribution of the participation ratio (top), density of states (middle), and relation between participation ratio and eigenvalue (bottom) for the 2D random-coupling model. The left graphs show the data for different power law exponents $s$, the right graphs show the data for different system sizes, for a large and a small values of $s$.} \label{Uebersicht/OVV_RCM_tav_4_2D}
\end{figure}

The random-coupling model in two dimensions confirms our expectation that finite-size effects are considerably weaker when $s$ is larger. The probability distribution of the participation ratio is essentially independent of the system size for $s=0.75$ for the range of system sizes shown in Figure \ref{Uebersicht/OVV_RCM_tav_4_2D}, top right. As for the diffusion model, the participation numbers decrease and the relaxation times increase as $s$ becomes larger. For $s$ close to 1, the density of states appears to follow a power law close to $E^{-s}$, which means that relaxation times are dominated by the small couplings, see middle row of  Figure \ref{Uebersicht/OVV_RCM_tav_4_2D}. This dominance of small couplings is also apparent from the bottom right graph for $s=0.75$, where a larger system size leads to smaller $E$ values, which can be explained by the smallest couplings becoming smaller for larger systems. On the other hand, the data for $s=0.25$ show that for a given energy the participation ratio increases with system size, indicating considerable finite-size effects, as is also visible from the top right graph. 

\section{Discussion and conclusion}

The investigation of tight-binding models with power-law distributed bond disorder $f(t) \sim t^{-s}$ has revealed several systematic trends as $s$ is increased towards 1: When $s$ is larger, there are more weak links, increasing effectively the strength of disorder. As a result, characteristic time scales become slower, localization becomes stronger, and finite-size effects are reduced. There is an increasing number of slowly relaxing eigenmodes localized on one or two lattice sites.  
In the one-dimensional diffusion model, the density of states and the relation between the participation ratio and energy (or, equivalently, relaxation time) follow nonuniversal power laws with exponents that depend on $s$. This is a generalization of the phenomenon of anomalous diffusion reported in such systems before \cite{Alexander1981}. We were able to explain these results in an intuitive way by using phenomenological considerations. Interestingly, the probability distribution of the participation ratio remains universal and follows the power law $\sim P^{-2}$ reported before for a constant distribution of couplings \cite{schaeferScalingBehaviourLocalised2024}. We provided a simple explanation also for this observation. 

In two dimensions, weak couplings do not interrupt transport as they do in one dimension, and therefore the long-wavelength modes of the diffusion model show the constant density of states known from the system without disorder. The participation ratio shows the power law distributions $\sim P^{-1.5}$  reported before for the case of a constant distribution of couplings \cite{schaefer2026different}. However, with larger $s$, these data become less visible due to an increasing number of strongly localized but slowly relaxing eigenmodes. 
We have shown that for $s>0.75$ the modes with the slowest relaxation times (i.e. smallest $E$) are modes localized on one site that is surrounded by four weak couplings. 

The two models studied in this paper have a wide range of applications: The diffusion model applies to all types of diffusion processes in disordered systems where the transported quantity is conserved, from heat transport to chemical diffusion and dispersal of animals between habitats. It also applies to oscillating systems consisting of elastically coupled units. Furthermore, the diffusion model applies also to nonlinear models when they are linearized around a homogeneous stationary state, as is done in systems with diffusion-driven instabilities. The eigenmodes of these models represent modes of relaxation towards the constant equilibrium solution \cite{brechtelMasterStabilityFunctions2018}. With increasing power-law exponent $s$, such relaxation processes become increasingly dominated by strongly localized modes. 

The random-coupling model applies to quantum-mechanical tight-binding systems where disorder resides in the couplings (i.e., hopping terms) and not in the on-site potentials. It also results from classical nonlinear dynamics of the SIS model for disease spreading when the equations are linearized around the non-infected state 
\cite{pastorsatorrasEpidemicProcessesComplex2015a,pastorsatorrasEigenvectorLocalizationReal2018}. 

In contrast to these two models, the original Anderson model behaves differently when disorder follows a power-law distribution. In the Anderson model, the couplings are all identical, and disorder resides in the on-site potentials. If these potentials become power-law distributed according to $f(\epsilon) \sim \epsilon^{-s}$ with $s \in [0,1]$ and an upper cutoff $\epsilon_{\mathrm{max}}$, there is an increasing number of sites with similar values of the on-site potential, leading to reduced disorder. We verified this in our numerical calculations, but decided not to show the data since this effect is pretty obvious.

\printbibliography
\end{document}